\documentclass{natureprintstyle}
\usepackage{epsfig,caption}
\usepackage{color}
\usepackage{bm}
\usepackage{graphicx}
\usepackage{longtable}
\usepackage{amssymb}
\usepackage{rotating,xcolor}
\usepackage{hyperref}
\hypersetup{
    hidelinks,
    colorlinks=false,
    urlcolor=black
}
\usepackage{multibib}
\newcites{meth}{Methods References}

\title{A ``Black Hole Star" Reveals the Remarkable Gas-Enshrouded Hearts of the Little Red Dots}
\author{Rohan P. Naidu\textsuperscript{1, 2, $\dagger$},
Jorryt Matthee\textsuperscript{3},
Harley Katz\textsuperscript{4},
Anna de Graaff\textsuperscript{5},
Pascal Oesch\textsuperscript{6,7,8},
Aaron Smith\textsuperscript{9},
Jenny E. Greene\textsuperscript{10},
Gabriel Brammer\textsuperscript{7,8},
Andrea Weibel\textsuperscript{6},
Raphael Hviding\textsuperscript{5},
John Chisholm\textsuperscript{11},
Ivo Labb\'e\textsuperscript{12},
Robert A. Simcoe\textsuperscript{1},
Callum Witten\textsuperscript{6},
Hakim Atek\textsuperscript{13},
Josephine F. W. Baggen\textsuperscript{14},
Sirio Belli\textsuperscript{15},
Rachel Bezanson\textsuperscript{16},
Leindert A. Boogaard\textsuperscript{17},
Sownak Bose\textsuperscript{18},
Alba Covelo-Paz\textsuperscript{6},
Pratika Dayal\textsuperscript{19},
Yoshinobu Fudamoto\textsuperscript{20},
Lukas J. Furtak\textsuperscript{21},
Emma Giovinazzo\textsuperscript{6},
Andy Goulding\textsuperscript{10},
Max Gronke\textsuperscript{22},
Kasper E. Heintz\textsuperscript{6,7,8},
Michaela Hirschmann\textsuperscript{23},
Garth Illingworth\textsuperscript{24},
Akio K. Inoue\textsuperscript{25,26},
Benjamin D. Johnson\textsuperscript{27},
Joel Leja\textsuperscript{28,29,30},
Kate Leonova\textsuperscript{31},
Ian McConachie\textsuperscript{32},
Michael V Maseda\textsuperscript{32},
Priyamvada Natarajan\textsuperscript{14,33,34},
Erica Nelson\textsuperscript{35},
David J. Setton\textsuperscript{10,36},
Irene Shivaei\textsuperscript{37},
David Sobral\textsuperscript{38,39},
Mauro Stefanon\textsuperscript{40,41},
Sandro Tacchella\textsuperscript{42,43},
Sune Toft\textsuperscript{7,8},
Alberto Torralba\textsuperscript{3},
Pieter van Dokkum\textsuperscript{33},
Arjen van der Wel\textsuperscript{44},
Marta Volonteri\textsuperscript{13},
Fabian Walter\textsuperscript{5},
Bingjie Wang\textsuperscript{28,29,30},
Darach Watson\textsuperscript{7,8}
}

\begin{document}
\maketitle
\let\thefootnote\relax\footnote{

\begin{affiliations}
\item MIT Kavli Institute for Astrophysics and Space Research, 70 Vassar Street, Cambridge, MA 02139, USA
\item NASA Hubble Fellow
$\dagger$ \href{mailto:rnaidu@mit.edu}{rnaidu@mit.edu}
All author affiliations listed at end of paper.
\end{affiliations}
}

\vspace{-3.5mm}
\begin{abstract}
The physical processes that led to the formation of billion solar mass black holes within the first 700 million years of cosmic time remain a puzzle\cite{Banados18,Yang20Poniuaena,Wang21quasar}. Several theoretical scenarios have been proposed to seed and rapidly grow black holes\cite{Woods19,Smith19SMBH,Inayoshi20,Volonteri21}, but direct observations of these mechanisms remain elusive. Here we present a source 660 million years after the Big Bang that displays singular properties: among the largest Hydrogen Balmer breaks reported at any redshift, broad multi-peaked H$\beta$ emission, and Balmer line absorption in multiple transitions. We model this source as a  “black hole star" (BH*) where the Balmer break and absorption features are a result of extremely dense, turbulent gas forming a dust-free ``atmosphere" around a supermassive black hole\cite{Inayoshi24,Ji25BlackThunder}. This source may provide evidence of an early black hole embedded in dense gas -- a theoretical configuration proposed to rapidly grow black holes via super-Eddington accretion\cite{AlexanderNatarajan14,Begelman08quasistars,Volonteri10,Schleicher13,CoughlinBegelman24}. Radiation from the BH* appears to dominate almost all observed light, leaving limited room for contribution from its host galaxy. We demonstrate that the recently discovered “Little Red Dots" (LRDs)\cite{Matthee24,Greene24,Kocevski24,Taylor24,Lin24} with perplexing spectral energy distributions\cite{Akins24,Wang24z8,Setton24b,Kokorev24,Labbe24Monster,Labbe25LRDs,Killi24} can be explained as BH*s embedded in relatively brighter host galaxies. This source provides evidence that SMBH masses in the LRDs may be over-estimated by orders of magnitude -- the BH* is effectively dust-free contrary to the steep dust corrections applied while modeling LRDs\cite{Greene24,Brooks24,Furtak24,Ji25BlackThunder}, and the physics that gives rise to the complex line shapes and luminosities may deviate from assumptions underlying standard scaling relations.
\end{abstract}

%%%%%%%%%%%%%%%%%%%%%%%%%%%%%%%%%%%%%%%%%%%%%%%%%%%%%

We recently observed `MoM-BH*-1' with JWST's NIRSpec instrument as part of the ``Mirage or Miracle" JWST program (GO-5224). MoM-BH*-1 was selected as a high priority target for spectroscopic follow-up based on its striking appearance in NIRCam images of the UDS extragalactic field\cite{Donnan24}. It stood out as the reddest source in this $\approx250$ arcmin$^{2}$ field (F277W-F356W$>2.5$ mag), appearing remarkably luminous (F444W$=25.4$ mag) and unresolved at $>3\mu$m, while apparently disappearing at shorter wavelengths (F200W $>28.5$ at 3$\sigma$; Fig. 1a). 

In Fig. 1b we show the 4.5h deep NIRSpec prism spectrum we obtained ($R\approx150$, $\approx1-5\mu$m; December 15th, 2024). In Fig. 1d we display a public archival 1.5h NIRSpec G395M spectrum taken by the EXCELS survey\cite{Carnall24} ($R\approx1500$, $\approx3-5\mu$m; December 19th, 2023). The redshift ($z_{\rm{spec}}=7.7569^{+0.0013}_{-0.0012}$) is confirmed from several features: a broad H$\beta$ emission line (FWHM$=3036^{+361}_{-506}$ km s$^{-1}$), H$\gamma$ absorption at the same redshift as deep H$\beta$ absorption, and a strong Balmer break between $\approx3-4\mu$m where the flux drops by a factor of $>20\times$ thereby explaining the extremely red NIRCam color ($f^{\rm{\nu}}_{\rm{F444W}}/f^{\rm{\nu}}_{\rm{F277W}}>20$). A narrow [OIII]$4960,5008$\AA\ doublet is detected ($3.5\sigma$) at a redshift consistent with the Balmer lines. 

\begin{figure*}
\includegraphics[width=\textwidth]{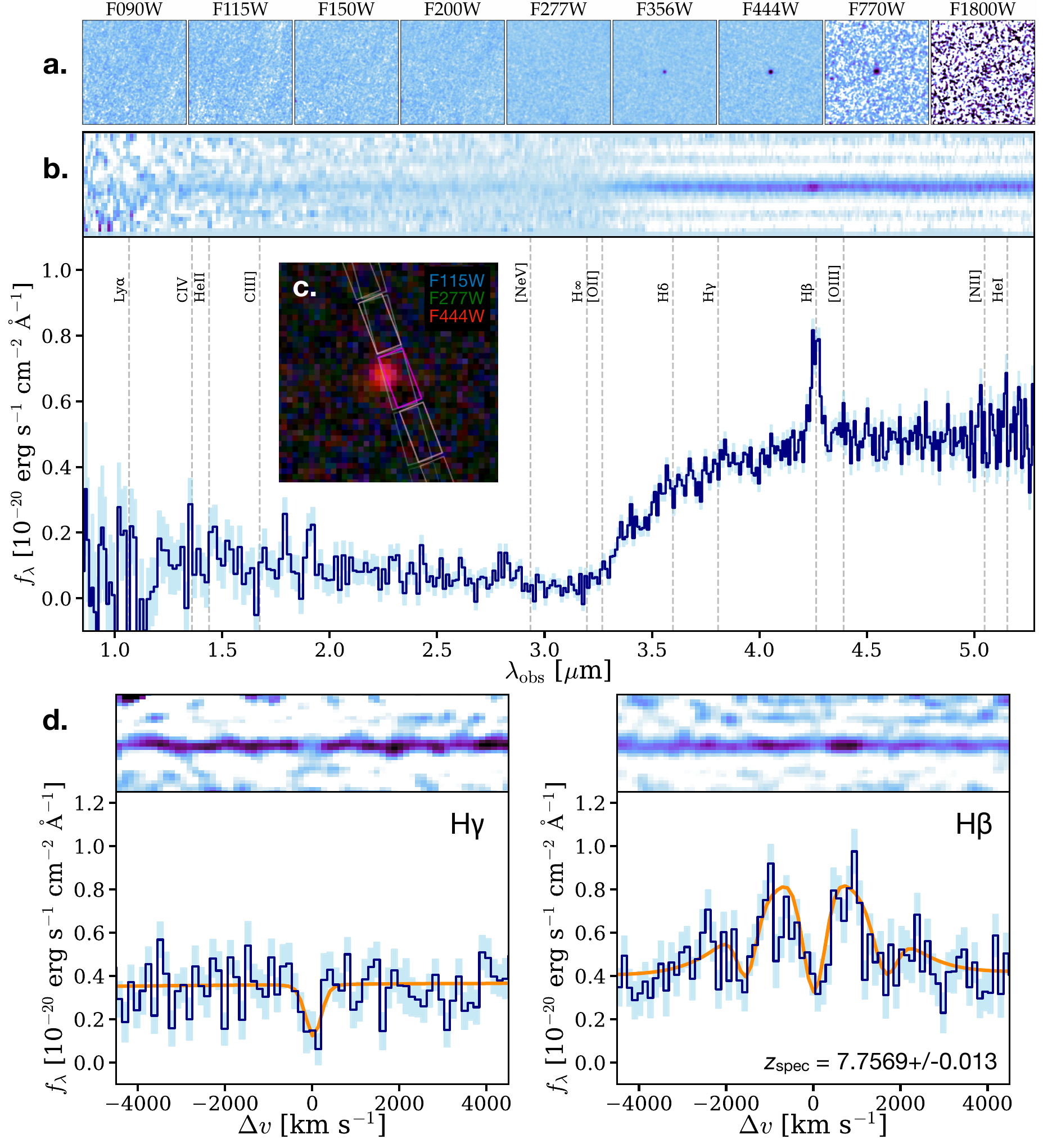}
\vspace{-4mm}
\caption{\textbf{JWST imaging and spectroscopy of MoM-BH*-1.} \textbf{Panel a.} $3\times3''$ NIRCam and MIRI images of MoM-BH*-1 spanning 0.9-18 $\mu$m. The source is point-like and detected ($>3\sigma$) only in the F356W, F444W, and F770W bands, apparently disappearing in the bluer bands. \textbf{Panel b.} The NIRSpec prism spectrum (navy blue) shows the disappearance is due to an enormous Balmer break. Key spectral features such as the Balmer series are marked with dashed lines. \textbf{Panel c.} Inset $1''$ RGB image shows the almost identical slit positions with which the source was observed with the prism (panel b) and G395M grating (panel d). \textbf{Panel d.} Deep absorption features in H$\gamma$ and H$\beta$ are evident in the G395M grating spectra. The location of the central absorption is consistent across both H$\beta$, H$\gamma$ as well as across the prism and grating spectra. The systemic redshift is based on the [OIII]$4960, 5008$\AA\ doublet. A representative draw from the emission line model posterior is plotted in dark orange (see Methods).
}
\vspace{-4mm}
\end{figure*}

\begin{figure*}[t!]
  \centering
  \includegraphics[width=\textwidth]{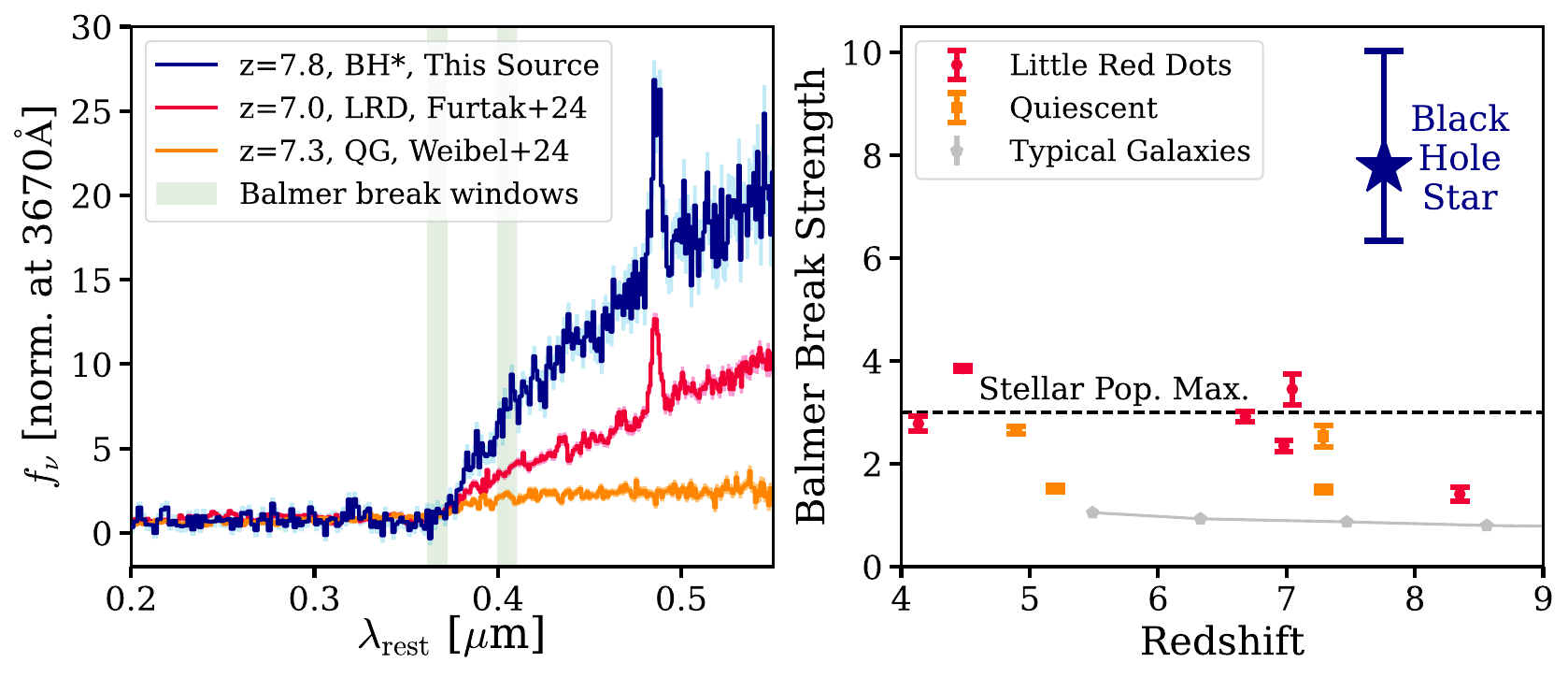}
  \caption{\textbf{The exceptional Balmer break strength of MoM-BH*-1.}  \textbf{Panel a.} Here we contrast MoM-BH*-1 against a quiescent galaxy\cite{Weibel24} and a Little Red Dot\cite{Furtak24,Furtak25,Ji25BlackThunder} that lie at a similar redshift ($z\approx7$) and display some of the strongest Balmer breaks reported yet ($\approx3$). The two wavelength windows we use to compute break strengths are highlighted in green -- these windows ([3620-3720]\AA\ and [4000-4100]\AA) are free of strong emission lines and are particularly suited for studying high redshift galaxies\cite{Wang24z8}. The spectra shown here are normalized in the blue window -- flux in this window is detected at $>4.5\sigma$ for MoM-BH*-1. \textbf{Panel b.} We compare break strengths of quiescent galaxies\cite{degraaff24,Weibel24QG,Strait23}, Little Red Dots with Balmer breaks\cite{Furtak24,Kokorev24LRDbbreak,Labbe24Monster,Wang24z8}, and stacks of star-forming galaxies\cite{Roberts-Borsani24} at similar redshifts as MoM-BH*-1. The dashed line represents the maximum break strength expected for a dust-free stellar population\cite{Wang24z8} and a Chabrier initial mass function\cite{Chabrier03}. MoM-BH*-1 displays the strongest Balmer break at these redshifts and lies well beyond this stellar population maximum.}
  \vspace{-6mm}
\end{figure*}

\begin{figure}[t]
  \centering
\includegraphics[width=\linewidth]{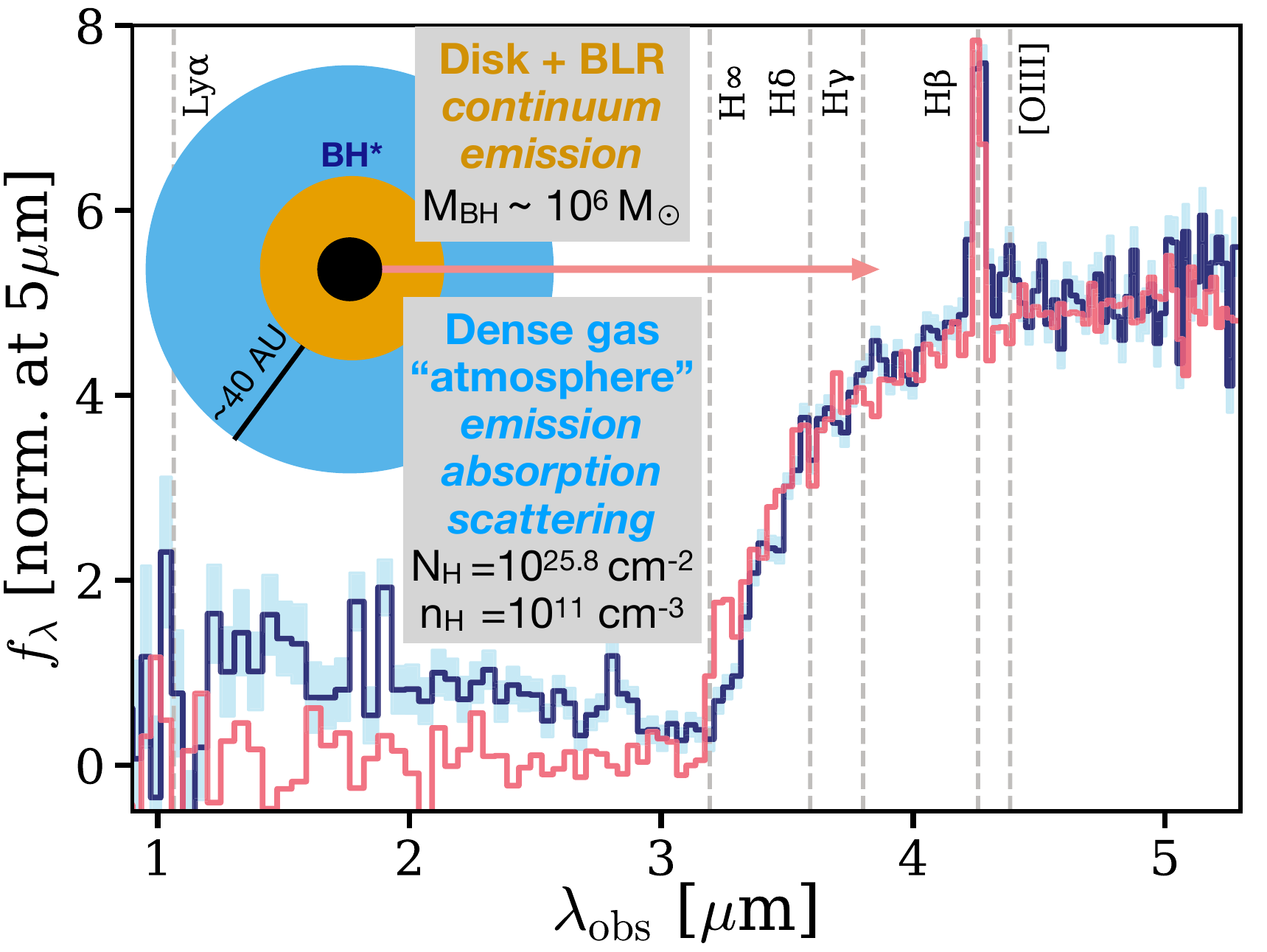}
  \caption{\textbf{Comparison against a mock spectrum of a ``black hole star" model}. The schematic depicts an SMBH within a $\approx40$ AU ``atmosphere" of dense gas -- the continuum is produced in hot regions close to the SMBH whereas absorption, scattering, and further emission occur in the dense gas atmosphere. The data (navy blue) are binned ($3\times$) to emphasize the continuum shape that our fiducial model (pink, with noise as per error spectrum) provides an excellent match to. The model is selected to reproduce the Balmer break strength and Balmer line EWs, while also matching the UV-faintness and MIRI long-wavelength data without having to invoke different mechanisms for lines and continuum. The narrow [OIII] emission and additional UV luminosity plausibly arise from the faint host galaxy, and are not captured by the BH* model (see Fig. 4). The excess flux around H$\infty$ is a \texttt{Cloudy} artifact due to modeling with a finite number of Hydrogen levels\cite{Ji25BlackThunder}.} 
  \vspace{-6mm}
\end{figure}

\begin{figure*}[t!]
  \centering
\includegraphics[width=0.95\textwidth]{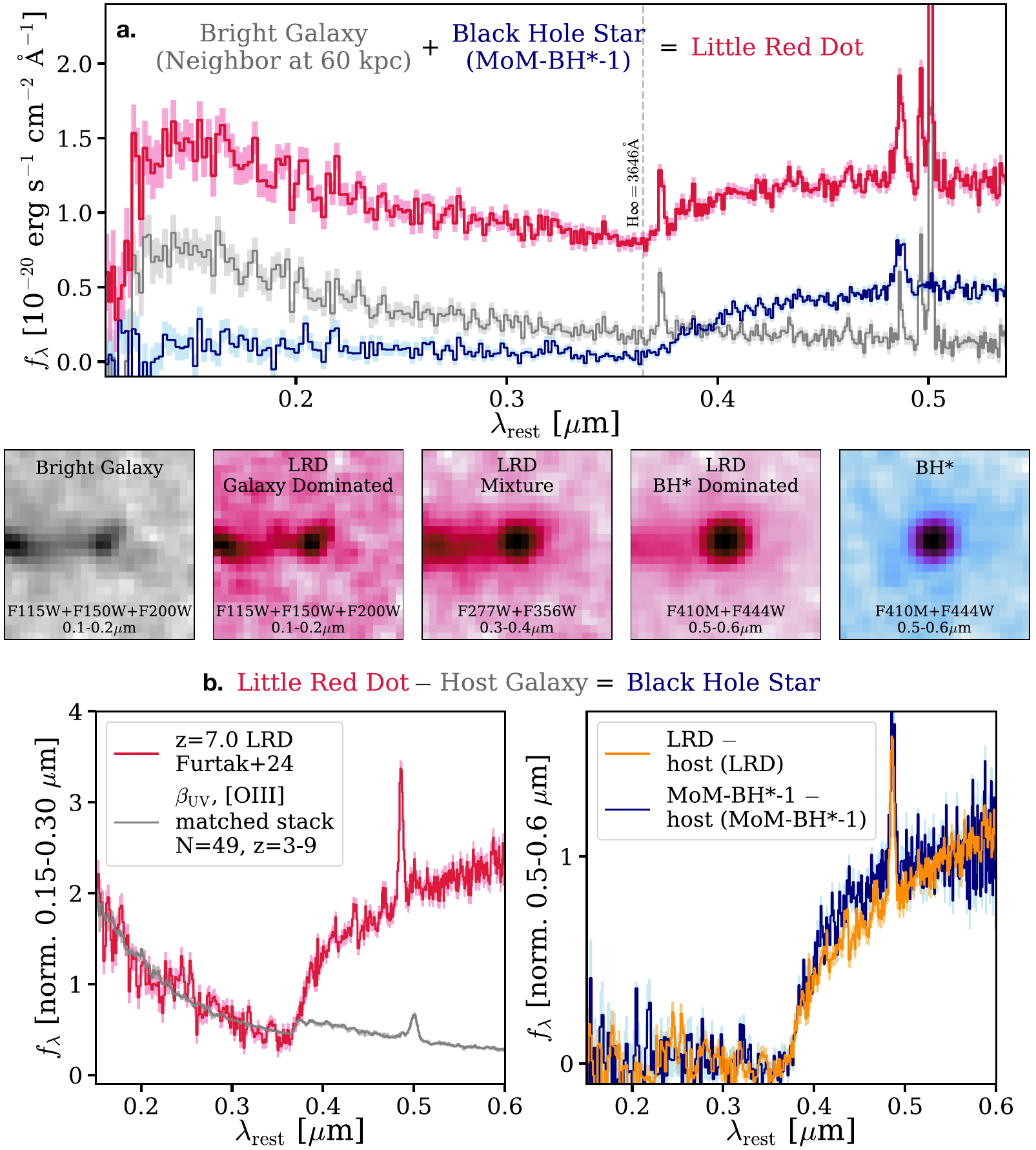}
  \caption{\textbf{Examples to illustrate Little Red Dots can be explained as BH*s embedded in comparably bright host galaxies}. \textbf{Panel a.} MoM-BH*-1 (blue) lies close to a $M_{\rm{\star}}\approx10^{9.5} M_{\rm{\odot}}$ galaxy at the same redshift (silver). These sources are expected to merge in $\approx100$ Myrs\cite{Puskas25}, and their superimposed spectrum (offset for clarity) and photometry ($1^{\prime\prime}\times1^{\prime\prime}$ NIRCam stamps) is shown in red. The combination bears a striking resemblance to the typical LRD -- a V-shaped SED, an inflection around $\rm{H}_{\infty}=3646\AA$, compactness in the rest-optical, a complex H$\beta$ profile with a broad component, and extended structure in the rest-UV. While the galaxy is dominant in the rest-UV, the BH* outshines it towards the rest-optical. \textbf{Panel b.} When the SED of a host galaxy is carefully subtracted from an LRD, the residual (orange) bears a striking resemblance to a BH*. Here, the host galaxy SED (silver) is a stack from the DAWN JWST archive selected to match the UV slope and weak [OIII] flux of this well-studied LRD\cite{Furtak24} (red). Interestingly, the stack resembles a ``mini-quenched" galaxy\cite{Strait23,Looser24,Witten24}, highlighting a possible connection between the evolution of supermassive black holes and their hosts. For the comparison in the right panel we similarly subtract a separate stack matched to MoM-BH*-1 (blue) to account for its very faint host galaxy.}
  \vspace{-4mm}
\end{figure*}

The strength of the Balmer break is remarkable. In Fig. 2 we compare MoM-BH*-1 with objects displaying Balmer breaks at similar redshifts: quiescent galaxies and a compilation of ``Little Red Dots" (LRDs; compact, red objects with broad Balmer lines). The maximum break strength expected for a dust-free stellar population assuming a typical IMF\cite{Chabrier03} is $\approx3$\cite{Wang24z8, Kriek06theory}. As an extreme case, a population comprised purely of A-type stars with the strongest breaks would have a strength $<5$\cite{Pickles98,MILES2011}. Crucially, MoM-BH*-1 is the only source that lies firmly beyond these limits with a break strength of $7.7^{+2.3}_{-1.4}$. A range of puzzling objects with a combination of broad Balmer lines and Balmer breaks have been discovered with JWST\cite{Furtak24,Wang24z8,Labbe24Monster}. However, crucially, all these sources fall around or below the observed and theoretical maximums for stellar populations, thereby permitting a wide variety of interpretations ranging from pure stars to pure AGN, combinations thereof, as well as more exotic explanations\cite{Wang24z8,Baggen24,Ji25BlackThunder,Ma24404,Bellovary25}. In contrast, it seems a relatively inescapable conclusion that the spectrum of MoM-BH*-1 does not arise from a stellar population.
 
The key to unraveling the break is the intense absorption in the Balmer lines (H$\beta$, H$\gamma$), which occurs simultaneously with H$\beta$ emission. Absorption in these non-resonant emission lines implies extreme gas densities ($n_{\rm{H}}\gtrsim10^{9}$ cm$^{-3}$) such that hydrogen atoms with populated $n=2$ shells are abundant\cite{Hall07,Inayoshi24}. Prior to JWST, this phenomenon was observed only in a handful of sources hosting supermassive black holes \cite{Aoki06,Hall07,Schulze18}, but it is now witnessed frequently in LRDs \cite{Matthee24,Labbe24Monster,Lin24,Juodzbalis24}. The absorption in MoM-BH*-1 is qualitatively similar to these LRDs, but is particularly strong, with missing flux at the line-center and over a broad velocity range.

The point-source morphology ($<100$ parsecs in F356W, 95$\%$ upper-limit) and extremely broad H$\beta$ emission (comparable to luminous $z>6$ quasars\cite{Yang23ASPIRE,Yue24EIGER}) further imply a supermassive black hole may be powering this source. Tentative detections of narrow forbidden lines (the [OIII] doublet) and a hint of variability ($30^{+7}_{-7}\%$ brightening in 56 days at $3-5\mu$m, albeit measured with different instruments; see Extended Data Fig. 5) add further evidence for a black hole. Such an extreme H$\beta/[\rm{OIII}]5008\AA$ ratio $>10$ ($11.4^{+4.2}_{-2.5}$), consistent with high gas densities where [OIII] is suppressed by collisional de-excitation ($\gtrsim10^6\ \rm{cm}^{-3}$), has been reported only in a single source at $z>6$, which appears to be a broad-line, variable AGN \cite{Furtak24,Furtak25,Ji25BlackThunder}. However, no known object, AGN or not, displays the singular Balmer break we report in this source.

Motivated by the evidence for a supermassive black hole and the strong Balmer absorption, following \cite{Inayoshi24,Ji25BlackThunder} we construct a grid of \texttt{Cloudy}\cite{Chatzikos23} spectral synthesis models where we embed a classical AGN accretion disk\cite{Williams87} within extremely dense gas (see Methods for details). Informed by the width of the H$\gamma$ and H$\beta$ absorption (FWHM$\approx300-500$ km s$^{-1}$), we model the absorbing gas with turbulent velocity. 

With this simple, idealized model we are able to match key features in this source including a deep, smooth Balmer break. In Fig. 3 we display our fiducial model selected from a grid of close to a million models, and spanning the extreme parameters demanded by this source. The model is selected to reproduce the EWs of H$\beta$ and H$\gamma$, the Balmer break strength, the UV weakness ($M_{\rm{UV}}>-18.5$), and the shape of the continuum out to $\lambda_{\rm{obs}}\approx20\mu$m constrained by MIRI. Extremely dense gas ($n_{\rm{H}}=10^{11}$ cm$^{-3}$, $N_{\rm{H}}=10^{25.8}$ cm$^{-2}$) with a high turbulent velocity (500 km s$^{-1}$) is necessary to produce these features. This velocity happens to match the width of the central absorption in H$\beta$. An important facet of this model compared to prior efforts\cite{Ji25BlackThunder} is that negligible dust attenuation ($A_{\rm{V}}=0.15$ mag vs. e.g., $A_{\rm{V}}>2$ mag) is invoked to match the continuum shape including the MIRI detection. This feature is consistent with stringent IR constraints ruling out significant dust in the LRDs \cite{Williams24,Setton25,Xiao25}. We emphasize that this modeling exercise is highly simplistic (e.g., the intrinsic AGN SED may be vastly different than assumed\cite{Sirko03} or the structure may be convective\cite{CoughlinBegelman24}), and only serves to provide broad physical intuition that dense gas enveloping a central engine may account for the singular observed features.

The detailed structure of the emission lines holds critical clues to the physical picture as well. In particular, the H$\beta$ line profile is remarkably symmetric with peaks and troughs mirrored on either side of the systemic redshift (Fig. 1, Extended Data Fig. 2). This makes it unlikely that random absorbers along the line of sight or inflows/outflows are responsible for the line structure. Instead, a coherent, symmetric absorption structure (such as e.g., a shell of gas) close to the object (as the central absorber is at the systemic velocity $42^{+80}_{-200}$ km s$^{-1}$) is our preferred solution for these features. In the Methods section we present a simple speculative model for this symmetry as arising from multiple scatterings of H$\beta$ that behaves like Ly$\alpha$.

The entire physical scenario presented here may be summarized as a ``Black Hole Star" (BH*). In a much more extreme avatar of the Balmer breaks observed in galaxies due to absorption in stellar atmospheres\cite{Kriek06,degraaff24,Weibel24QG}, here we have a black hole seen through a dense, turbulent envelope of Compton thick gas spanning $\approx10-100$ AU. The resulting SED has the characteristic features of an SMBH such as broad-lines, but also features that are traditionally associated with evolved stellar populations such as a Balmer break. 

To confront the puzzle of $\approx10^{9} M_{\rm{\odot}}$ SMBHs that are already in place by $z\gtrsim7.5$\cite{Banados18,Yang20Poniuaena,Wang21quasar}, theories of SMBH growth have envisioned channels of seeding massive BHs as well as growing them at a rapid, super-Eddington pace. As an early, growing black hole ($M_{\rm{BH}}\approx10^{6-7} M_{\rm{\odot}}$; see Methods), MoM-BH*-1 displays key features predicted by these models. For example, a class of models\cite{AlexanderNatarajan14,CoughlinBegelman24,Volonteri10} predicts that if a black hole is ensconced in dense gas (e.g., in a dense gas-rich nuclear star cluster or in a quasi-spherical gas distribution), the high opacity can trap the accretion radiation or ``convect" it away such that gravity may overcome radiative feedback and exceed the Eddington limit. MoM-BH*-1 may be undergoing an active super-Eddington burst, or perhaps it is in the end stages of such an episode wherein the gas envelope that has nourished it is still in place. Naively applying local scaling relations\cite{Vestergaard06} under standard assumptions to this source indicates the latter scenario ($L/L_{\rm{Edd.}}=18^{+7}_{-3}\%$), but accounting for the extreme conditions may favor the former ($L/L_{\rm{Edd}}\approx5-10$; see Methods).

It is separately noteworthy that the faint host galaxy surrounding the BH* is a low-mass (and hence, perhaps, metal-poor) dwarf galaxy ($M_{\rm{*}}<10^{8.5}M_{\rm{\odot}}$). We derive this limit based on the observed UV luminosity that we scale empirically using a large spectroscopic reference sample of dwarf galaxies\cite{Naidu24}. The UV is likely some mixture of BH* and galaxy, hence this is an upper limit -- but this is the wavelength where the galaxy shines the brightest relative to the BH*\cite{Matthee24ALTclustering}. Intriguingly, this low-mass galaxy appears to be associated to a more massive ($\approx10^{9.5} M_{\rm{\odot}}$) spectroscopically confirmed galaxy at a projected distance of only $\approx60$ proper kpc and $\Delta z<0.01$ (Fig. 4). Theories of ``direct collapse black holes" that form out of primordial gas predict such proximity to an ionizing source may be the key to suppressing the formation of molecular hydrogen thereby aiding direct collapse\cite{Agarwal16,Natarajan17} -- if more BH*s are found in similar configurations, this may be a telling sign.

How common is the BH* phase of SMBH growth? In Fig. 4 we argue the numerous ``Little Red Dots" recently revealed by JWST may be hosting BH*s\cite{Matthee24,Greene24,Kocevski24,Taylor24,Lin24}. Explaining the unique constellation of features found in these sources has proven challenging -- they display V-shaped SEDs that rise into the rest-UV as well as in the rest-optical, in some cases Balmer breaks, and generally X-ray and IR weakness with a lack of UV-variability but hints of optical variability \cite{Akins24,Wang24z8,Setton24b,Labbe24Monster,Labbe25LRDs,Furtak25,Ji25BlackThunder}. There are puzzling correlations as well: the broad-to-total H$\alpha$ line flux appears to closely track the UV-to-optical strength\cite{Matthee24}. No known observed combination of star-forming galaxies and AGN can account for all these properties self-consistently\cite{Ma24404,Labbe24Monster}.

Here we illustrate that treating MoM-BH*-1 as an effectively pure template for the AGN component of LRDs and combining it with a star-forming galaxy of matched UV-brightness accounts for the key perplexing LRD properties. In our proposed composite picture, the star-forming galaxy dominates in the UV (extended morphology, narrow Ly$\alpha$ line)\cite{Volonteri25}, whereas the BH* dominates in the rest-optical (Balmer break, broad Balmer lines, variability expected from AGN around H$\alpha$). X-ray weakness may be understood as a consequence of the Compton thick gas envelope\cite{Maiolino24Chandra} while the FIR-weakness is a consequence of much lower dust attenuation than all LRD studies assume since the BH* SED is intrinsically UV-weak below the Balmer break. We demonstrate this composite solution using fully empirical examples in Fig. 4 -- we show the combination of MoM-BH*-1 with a neighboring star-forming galaxy produces a source satisfying LRD color and morphology selection criteria\cite{Kokorev24,Labbe25LRDs} as well as subtle features such as a Balmer break and inflection around H$\infty$\cite{Setton24b}. We also show that when an appropriate host galaxy SED is subtracted from a well-studied LRD\cite{Furtak24}, the residual bears a striking resemblance to a BH* (Fig. 4b). The diversity in the relative contribution and properties of each component (e.g., star-formation history of the host, gas density around the BH*) may account for the full diversity seen in LRD properties. For example, objects like A2744-45924\cite{Labbe24Monster} may host BH*s that contribute all the way into the UV as evidenced by FeII emission-line features (possibly due to a lower covering fraction or less dense gas), whereas objects with very strong narrow H$\alpha$ emission on top of the broad lines\cite{Matthee24} may be dominated by the host galaxy even in the rest-optical. These observations provide a possibly complete, fully empirical solution to the perplexing LRD characteristics: varying contributions of BH* and star-forming galaxies can reasonably reproduce all of their observed properties. 

\captionsetup[figure]{labelformat=empty}
\renewcommand{\thefigure}{Extended Data \arabic{figure}}
\renewcommand{\thetable}{Extended Data \arabic{table}}

\begin{center}
{\bf \Large \uppercase{Methods} }
\end{center}

\begin{table}
\centering
\caption{Summary of key properties.}
\begin{tabular}{lr}
\hline
Property & Measurement \\
\hline
R.A. [deg] & 34.3974841\\
Dec. [deg] & $-5.1351163$\\
Redshift & $7.7569^{+0.0013}_{-0.0012}$\\
Break Strength ($f^{\nu}_{4050\rm \AA}/f^{\nu}_{3670\rm \AA}$) & $7.7^{+2.3}_{-1.4}$\\
$f^{\nu}_{4050 \rm \AA}$ [nJy] & $140^{+5}_{-5}$\\
$f^{\nu}_{3670 \rm \AA}$ [nJy] & $18^{+4}_{-4}$\\
Size in F356W ($99\%$ limit) [pc] & $<117$ \\
$M_{\rm{UV}}$ & $-18.1^{+0.2}_{-0.2}$\\
UV slope ($\beta_{\rm{UV}}$) & $-0.8^{+2.0}_{-0.2}$\\
Variability amplitude (56 days, $3-5\mu$m) & $30^{+7}_{-7}\%$\\
Host Galaxy Mass (from $M_{\rm{UV}}$, $95\%$ limit) & $<10^{8.5} M_{\rm{\odot}}$\\
Host Galaxy Mass (from SED fitting) & $<10^{8.8} M_{\rm{\odot}}$\\
\hline
\end{tabular}
\end{table}

\begin{table}
\centering
\caption{PSF-matched photometry\cite{Weibel24}. \label{table:photom}}
\begin{tabular}{lr}
\hline
Band & Flux (nJy) \\
\hline
\multicolumn{2}{c}{HST (ACS, WFC3)} \\
\hline
F435W & $-13\pm13$ \\
F606W & $-2\pm8$ \\
F814W & $3\pm8$ \\
F125W & $-2\pm17$ \\
F160W & $-3\pm15$ \\
\hline
\multicolumn{2}{c}{JWST (NIRCam, MIRI)} \\
\hline
F090W & $11\pm8$ \\
F115W & $15\pm8$ \\
F150W & $4\pm6$ \\
F200W & $5\pm5$ \\
F277W & $8\pm4$ \\
F356W & $93\pm5$ \\
F410M & $206\pm10$ \\
F444W & $243\pm12$ \\
F770W & $465\pm52$\\
F1800W & $1269\pm615$\\
\hline
\end{tabular}
\end{table}

\begin{table}[t]
\centering
\caption{Emission and absorption line measurements. All lines are constrained to the systemic redshift tied to [OIII], with the exception being the H$\beta$ absorbers whose velocity offset in km s$^{-1}$ is indicated in parentheses. Note the remarkably symmetric locations of the recovered H$\beta$ absorption.}
\begin{tabular}{lccc}
\hline
Line & Flux & EW & FWHM \\
 & [10$^{-20}\times$ & [\AA] & [km s$^{-1}$] \\
  & erg s$^{-1}$ cm$^{-2}$] & & \\
\hline
H$\gamma$ & $-9^{+3}_{-4}$ & $-3^{+1}_{-1}$ & $284^{+227}_{-181}$ \\
$\rm{[OIII]}$5008\AA\ & $12^{+3}_{-3}$ & $3^{+1}_{-1}$ & $135^{+175}_{-95}$\\
H$\beta$ & 201$^{+27}_{-24}$ & $57^{+8}_{-7}$ & $3036^{+361}_{-506}$\\
H$\beta$ ($+42^{+80}_{-200}$) & $-38^{+23}_{-14}$ & $-11^{+7}_{-4}$ & $571^{+97}_{-164}$\\
H$\beta$ ($+1556^{+232}_{-1378}$) & $-16^{+9}_{-15}$ & $-5^{+3}_{-4}$ & $505^{+137}_{-216}$\\
H$\beta$ ($-1532^{+345}_{-113}$) & $-15^{+7}_{-8}$ & $-4^{+2}_{-2}$ & $420^{+193}_{-221}$\\
\hline
\end{tabular}
\end{table}

\noindent
{\bf Observations and Data Reduction}
\\
\noindent
MoM-BH*-1 has been observed with JWST by three programs. It was imaged in Cycle 1 by the PRIMER survey\cite{Donnan24} (JWST-GO-1837; PI: J. Dunlop) using the MIRI (January 5th and 16th, 2023) and NIRCam (August 7th and 9th, 2023) instruments. In Cycle 2 (December 19th, 2023) the EXCELS survey\cite{Carnall24} (JWST-GO-3543) obtained 1.5h of NIRSpec G395M spectroscopy. In Cycle 3 (December 15th, 2024), we targeted MoM-BH*-1 as part of the ``Mirage or Miracle" NIRSpec prism survey (JWST-GO-5224). MoM was designed to collect a statistical sample of luminous $z>10$ galaxies and quantify the contamination fraction in high-redshift galaxy selections. For this purpose, MoM is surveying the wide-area JWST extragalactic legacy fields (COSMOS, UDS). We included MoM-BH*-1 as a high priority target second in importance only to luminous $z>10$ sources in our UDS masks because of its occurrence in multiple priority target lists -- AGN/LRD candidates selected based on compact morphology and template fitting with \texttt{EAZY}\cite{Brammer07}, extremely massive galaxy candidates based on SED fitting, and sources with peculiar red colors. Separately, this source had also been identified as a photometric LRD candidate based on a template fitting method\cite{Kocevski24}.

We use the v7.2 images of the PRIMER field released on the DAWN JWST archive (DJA) reduced using the \texttt{grizli} software. Details about the imaging data reduction may be found in \cite{Valentino23}. PSF-matched photometric catalogs based on these images were produced in \cite{Weibel24} that we deploy in this work. Fluxes for the source from these catalogs are listed in Table 2. We use the public v3 NIRSpec reductions of the EXCELS grating data from the DJA derived using the \texttt{msaexp} software, details of which can be found in \cite{msaexp,Heintz24,degraaff24}. The MoM data is reduced with the same pipeline following the exact same choices as the v3 DJA reductions.

\noindent
{\bf Emission Line Fitting}
\\
\noindent
We use a custom NIRSpec emission line fitting package (Hviding et al., in prep.) to simultaneously fit emission lines in the grating and prism spectra. The advantage of this approach is that despite the low-SNR in either mode, features may be robustly recovered due to their occurrence at the same wavelength across both dispersers. This package self-consistently accounts for under-estimated errors in the spectra by computing the scatter in the continuum, systematic flux offsets, wavelength offsets, and calibration uncertainty across both modes based on empirical data. The underlying LSF is that of an idealized point source\cite{degraaff24LSF}, but this LSF is rescaled by a nuisance parameter to account for calibration uncertainties\cite{Wang24}.

We first fit the H$\beta$ line and [OIII] doublet, and then use the redshift as a prior to fit H$\gamma$ (shown in Fig. 1). We model H$\beta$ as a single emission line with three absorbers constrained to have negative flux (one at line center, and two on either side of zero velocity) motivated by the symmetric absorption troughs on either side of the central double-peak (see Extended Data Fig. 2). The systemic redshift is tied to [OIII] and the broad H$\beta$ component, with the absorbers allowed to range freely. The number of absorbers is decided based on the maxima reached in the reduced $\chi^{2}$ which is similar for three and four absorbers, but we opt to exercise parsimony. See Extended Data Fig. 2 for further evidence that this large number of absorbers may be merited to account for secondary peaks at $\pm2400$ km s$^{-1}$. We found that including a narrow H$\beta$ emission line at the systemic redshift leads to a completely unconstrained flux degenerate with absorption and no improvement, so we neglect this component. The resulting fits, where we sample the posterior with the NUTS sampler implemented in \texttt{numpyro}\cite{numpyro} are shown in Extended Data Figure 1 and reported in Extended Data Table 3. 

While the formal errors on the derived fluxes and line-widths of many of the components is significant, the key features relevant to this analysis are robustly recovered -- extremely broad H$\beta$ emission, and deep, broad absorption wiping out $\approx25\%$ of the emission flux, including $\approx100\%$ of the flux at line-center. Another notable aspect of these fits is the location of the two absorbers -- they are recovered at very similar velocities, but on either side of line-center ($\approx\pm1500$ km s$^{-1}$), albeit with significant uncertainties ($-1532^{+345}_{-113}$ and $+1556^{+232}_{-1378}$ km s$^{-1}$). We explore this symmetry, which extends not only to the location of the absorbers but also to the detailed structure of the entire emission line profile over few 1000 km s$^{-1}$ in Extended Data Fig. 2.

\begin{figure}[t]
  \centering
\includegraphics[width=0.4\textwidth]{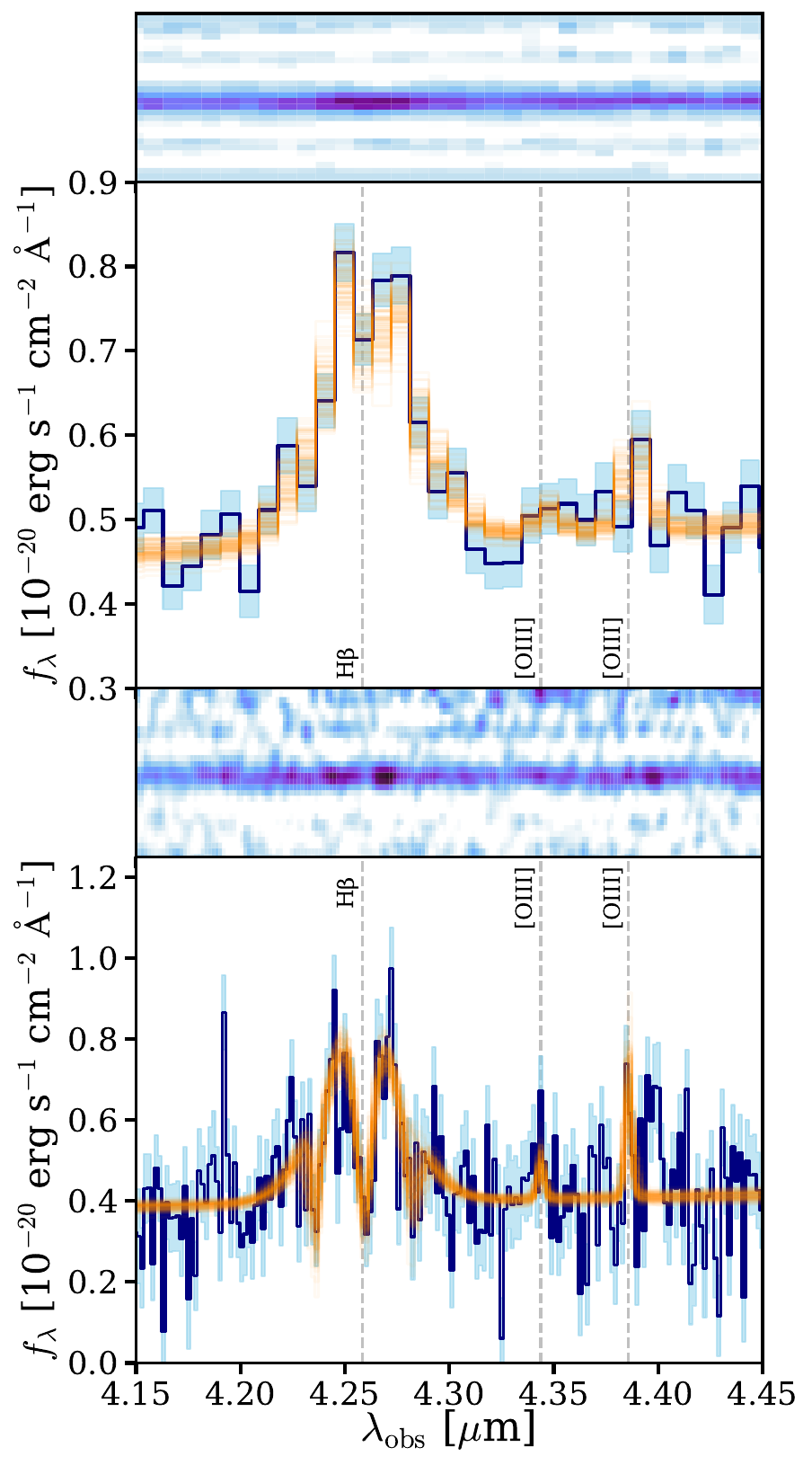}
  \caption{\textbf{Extended Data Figure 1 $|$ Simultaneous emission line fits to the prism (top) and grating (bottom) spectra.} 100 draws from the posterior are shown in orange. The consistency of features across both modes inspires confidence in their reality. For example, the detailed structure of the H$\beta$ line -- a central absorber, extremely broad wings, and absorption even in the wings -- recurs in both panels. Similarly, the existence of narrow [OIII]5008\AA\ emission would be difficult to discern in either mode by itself, but is recovered in the joint fit.} 
  \vspace{-4mm}
\end{figure}

\begin{figure}[ht!]
  \centering
\includegraphics[width=0.4\textwidth]{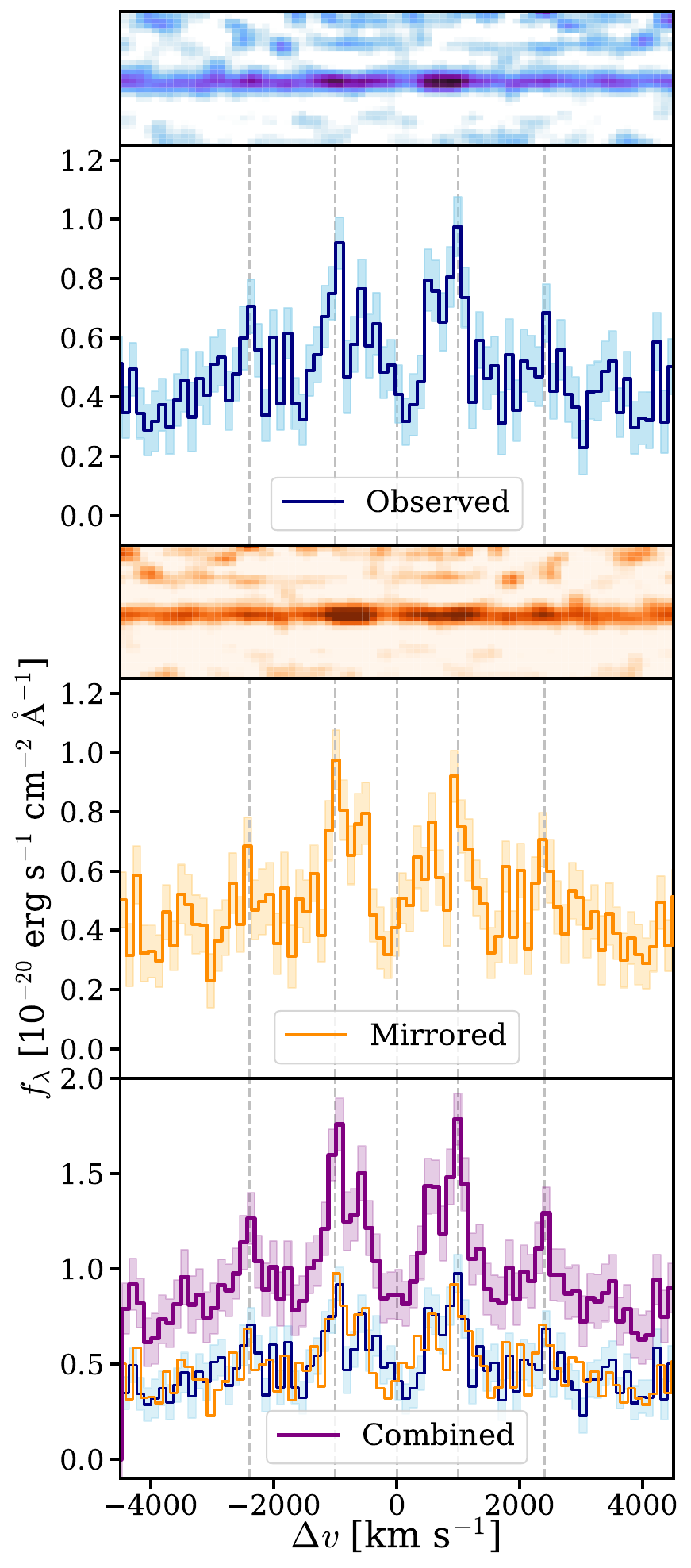}
  \caption{\textbf{Extended Data Figure 2 $|$ The remarkable symmetry of the H$\beta$ line profile points to a symmetric configuration of absorbing gas.} Here we compare the observed line profile (top, blue) to a line profile mirrored around the systemic redshift derived from [OIII] (center, orange). In the bottom panel we co-add these spectra. These profiles display peaks ($\pm1000$ km s$^{-1}$, and perhaps also $\pm2400$ km s$^{-1}$) and troughs ($\pm1500$ km s$^{-1}$; see absorber locations in Table 3) at similar velocities. This potential symmetry has important implications. First, it means the recovered systemic redshift from [OIII] is robust. Furthermore, these features are highly unlikely to be the result of random absorption systems or inflows/outflows that are fortuitously aligned at the same positive and negative velocities. Instead, this alignment suggests the presence of a symmetric absorbing structure (e.g., shells of gas).} 
  \vspace{-4mm}
\end{figure}

\noindent
{\bf Morphology}
\\
\noindent
We use \texttt{pysersic}\cite{Pasha23} to fit a Sersic profile to the imaging. We focus on the F356W and F444W imaging where the source is well-detected. We follow the exact same procedure as described in \cite{Weibel24QG} -- briefly, we build an empirical PSF using stars in the field, and then use this PSF with \texttt{pysersic} to sample the posterior Sersic parameters with \texttt{numpyro}. The source is unresolved, and we are able to place a $99\%$ upper limit on the effective radius of $<117$ pc consistent with the BH* interpretation. 

\noindent
{\bf \texttt{Cloudy} modeling}
\\
\noindent
That extremely dense gas might blanket the Little Red Dots was already inferred in the work that defined this class of sources\cite{Matthee24} -- $\approx10\%$ of these LRDs showed clear signs of Balmer absorption, with this being a lower limit on the incidence due to resolution and SNR considerations. Inspired by the Balmer absorption in LRDs, and using \texttt{Cloudy} models for continuum absorption, \cite{Inayoshi24} demonstrated that the very gas producing strong Balmer absorption likely also produces Balmer breaks. However, note the breaks in these models are weaker than what we require to explain MoM-BH*-1, and are abrupt instead of the smooth rollover seen in this source. Here we build on this pioneering work.

\begin{table}
\centering
\caption{Grid of \texttt{Cloudy} parameters.}
\begin{tabular}{lccc}
\hline
Property & Grid-Points \\
\hline
Temperature ($T_{\rm{BB}}$, [K]) & $10^{4}$, $5\times10^{4}$, $\mathbf{10^5}$, $5\times10^{5}$\\
Optical to X-ray index ($\alpha_{\rm{OX}}$) & $-1.0$, $\mathbf{-1.5}$, $-2.5$, $-5.0$\\
UV slope ($\alpha_{\rm{UV}}$) & $\mathbf{-0.1}$, $-1.0$, $-2.0$, $-2.5$\\
X-ray slope ($\alpha_{\rm{X}}$) & $-0.1$, $\mathbf{-0.5}$, $-1.5$, $-2.5$\\
Ionization Parameter $\log(\rm{U})$ & $\mathbf{-2.5}$, $-1.5$, $-0.5$\\
Gas Density ($n_{\rm{H}}$, [cm$^{-3}$]) & 9.0, 9.5, 10.0\\
 & 10.5, \textbf{11.0}, 11.5, 12.0\\
Column Density ($N_{\rm{H}}$, [cm$^{-2}$]) & 21.0, 22.0, 23.0, 23.5\\
& 24.0, 24.5, 25.0, \textbf{26.0}\\
Turbulence ($v_{\rm{turb.}}$, [km s$^{-1}$]) & 10, 50, 100, 200\\
& 300, 400, \textbf{500}\\
Metallicity ([Fe/H]) & $\mathbf{-2}$, $-1$, 0\\
Dust (post-processing, $A_{\rm{V}}$) & 0 to 3, \textbf{0.15} \\
\hline
\end{tabular}
\end{table}

The grid of parameters we explore is summarized in the Table above. This grid spans a wide range to capture the extreme spectral shape at hand. We start with an intrinsic AGN continuum SED that is parametrized using a series of power laws and with a ``big bump" temperature\cite{Williams87}. This AGN continuum is passed through gas surrounding the central source that is defined in terms of its gas density, column density, metallicity, and turbulent velocity. The irradiation of the cloud is modulated by an ionization parameter ($\log(\rm{U})$). A large turbulent velocity of the absorbing gas $\approx500$ km s$^{-1}$ is motivated by the significant width of the absorption lines we observe (see Fig. 1). This implies a very high Mach number which is consistent with recent models of AGN disks\cite{Hopkins24}. The turbulence is critical in producing a smooth rollover instead of a sharp break\cite{Ji25BlackThunder}. Finally, we apply a uniform dust screen\cite{Cardelli89} with $A_{\rm{V}}=0-3$ in the post-processing -- this is the only post-processing step we apply. 

To select models that are consistent with the data, we require the following conditions: (1) H$\beta$ in emission with $30<$EW [\AA]$<45$, (2) H$\gamma$ in absorption with $-5<$EW [\AA]$<0$ \AA, (3) a strong Balmer break and high optical-to-UV ratio such that $f^{\lambda}_{\rm{4.5\mu m}}/f^{\lambda}_{\rm{1.8\mu m}}>6$ and $f^{\lambda}_{\rm{4.5\mu m}}/f^{\lambda}_{\rm{2.8\mu m}}>3$, and (4) fluxes in MIRI bands within $2\sigma$ of the observations. The few thousand models that satisfy these constraints are re-simulated at higher resolution, retaining only Hydrogen that is relevant to the key features for simplicity and speed. Of these, we select the one that closely follows the detailed shape of the continuum (bolded parameters in Table). Additionally, we experiment with the covering factor, and distance between the gas and central source and find these quantities to be degenerate with the AGN SED and ionization parameter. We find the ``net transmitted" flux (i.e., the sum of the attenuated incident continuum and diffuse continuua/lines) produces a good match to the data, not the ``total" flux (which also includes reflected continuua/lines).

\begin{figure}[t]
  \centering
\includegraphics[width=0.49\textwidth]{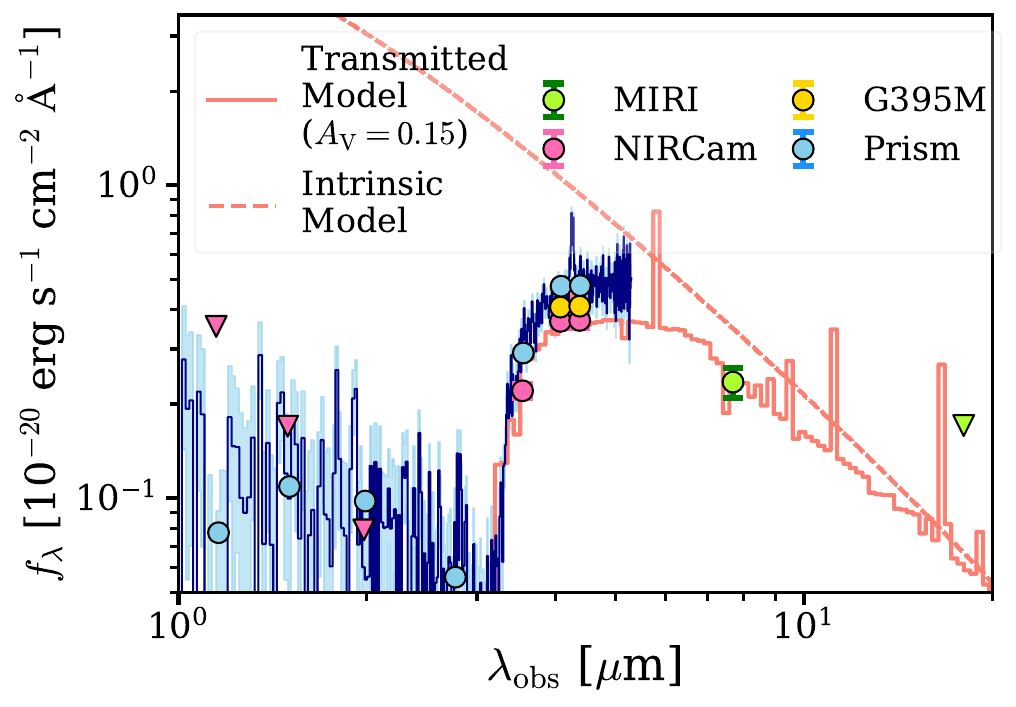}
  \caption{\textbf{Extended Data Figure 3 $|$ Comparison of fiducial model against long-wavelength data.} Not only is this virtually dust-free model ($A_{\rm{V}}=0.15$ mag) able to explain the strong Balmer break and UV-weakness, but also the ``turn-over" at infrared wavelengths traced by MIRI (shown in green) and generically reported for the LRDs\cite{Akins24,Williams24}. The model spectrum (salmon) is normalized to the NIRCam fluxes that are closest to the epoch of the MIRI observations. See text and Extended Data Fig. 5 for discussion of variability.} 
  \vspace{-4mm}
\end{figure}

\begin{figure*}
  \centering
\includegraphics[width=\textwidth]{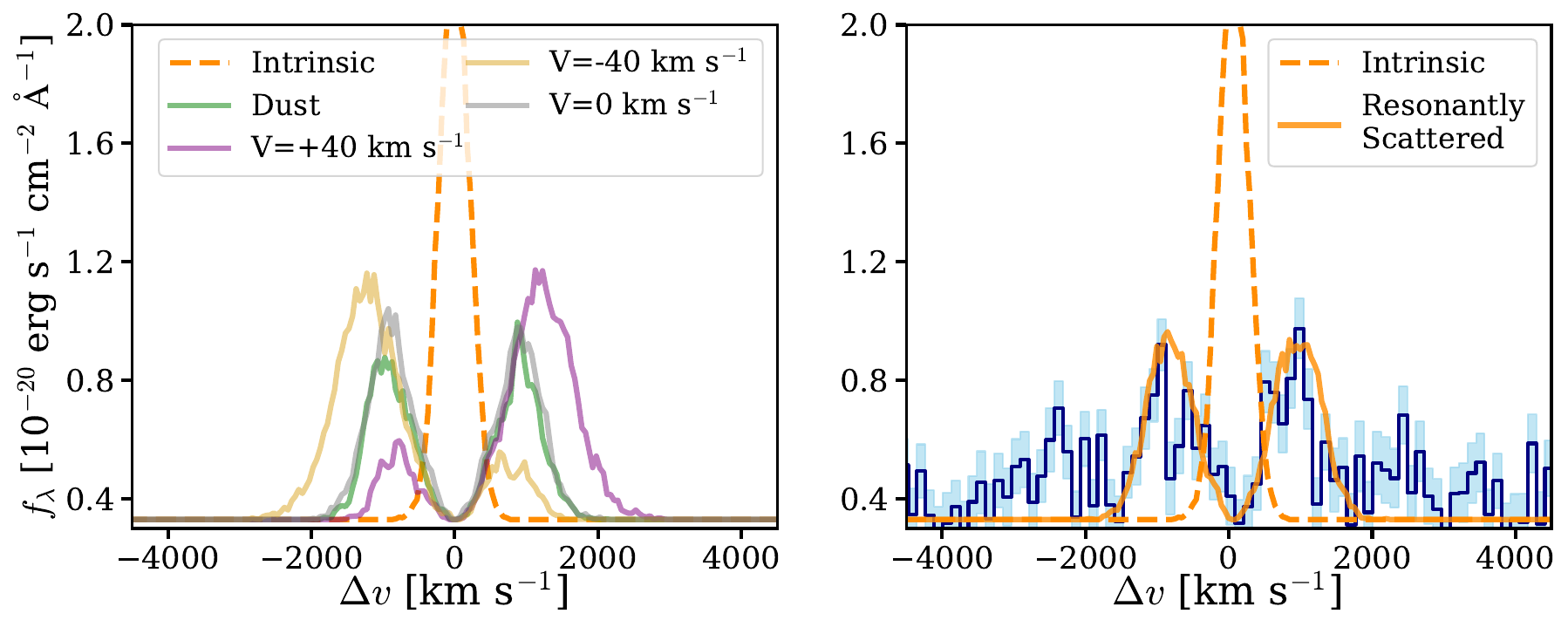}
  \caption{\textbf{Extended Data Figure 4 $|$ Resonant Balmer scattering in the BH*?} \textbf{Left:} Through simple radiative transfer calculations with a shell model we demonstrate that H$\beta$ behaves like Ly$\alpha$ if radiative decay transitions into the 2p state are suppressed. The trends that are well-known in the Ly$\alpha$ literature and relevant to our situation hold here. A narrow intrinsic line is scattered into double peaks. Dust weakens the blue peak preferentially, inflows/outflows boost one peak relative to another, and static shells produce equally strong peaks. \textbf{Right:} We model the two main peaks of the MoM-BH*-1 profile with a static, dust-free shell. This toy model does not capture the full extent of the broad wings which implies models with a more complicated geometry need to explored, especially with higher S/N data.} 
  \vspace{-4mm}
\end{figure*}

\noindent
\textbf{Key Features of Fiducial \texttt{Cloudy} Model: Blanketed by Gas, not by Dust}
\\
\noindent
First, we emphasize that this model (shown in Fig. 3 and Extended Data Fig. 3) is \textit{a} model that matches various features of interest based on our limited grid. The relevant parameter space that we have done our best to sample is high dimensional, degenerate, and much remains unknown (e.g., the intrinsic SEDs of early AGN). Therefore, we caution against detailed inferences beyond the feasibility of the broad physical picture  that we present here (an accretion disk embedded in dense gas).

Notably, the selected model features an extreme column density ($\approx10^{25.8}$ cm$^{-2}$) comparable to the most enshrouded systems observed\cite{Akylas24} and a high gas density ($10^{11}$ cm$^{-3}$) conducive to Balmer absorption. The high turbulent velocity of 500 km$s^{-1}$ is commensurate with the width of the central H$\beta$ absorber. Metal-poor gas expected of a dwarf galaxy at $z\approx8$ is preferred. The AGN slope parameters are well within the range of literature SEDs\cite{Pacucci24,Ji25BlackThunder}.

An important feature of the fiducial model is that it is virtually dust-free ($A_{\rm{V}}=0.15$ mag). The UV-weakness arises entirely due to the extreme Hydrogen opacity. This is a critical constraint on LRD models that typically invoke significant amounts of dust to suppress strong UV emission from classical AGN\cite{Labbe24Monster,Ma24404,Wang24} or to explain the weakness of H$\beta$ relative to H$\alpha$\cite{Brooks24}. While significant $A_{\rm{V}}\approx2-3$ helps dense gas AGN models produce a smooth Balmer break in the rest-optical\cite{Ji25BlackThunder}, this is ruled out by longer wavelength constraints in our source (see Extended Data Fig. 3) and more generally by stringent IR constraints on the LRDs\cite{Setton25,Xiao25}.

\noindent
{\bf Resonant Balmer Scattering? Insights from Ly$\alpha$-like Shell Models} 
\\
\noindent
The H$\beta$ profile of MoM-BH*-1 bears a remarkable resemblance to double-peaked Ly$\alpha$ \cite{Verhamme06,Gronke17,NM22}. This motivates us to explore whether H$\beta$ is indeed behaving like a resonant line under the high densities in the BH*. It may be the case that this is a common phenomena in the LRDs, but that a clearly symmetric double-peaked line with large separation is easier to observe in MoM-BH*-1 due to negligible H$\beta$ emission from the sub-dominant host galaxy. 

Although H$\beta$ normally has an easy cascade escape route, saturating the 2p state can effectively trap H$\beta$ photons, which may be plausible in a BH* through Ly$\alpha$ pumping in these extremely optically thick environments\cite{Smith17resonantscatter}. We test this idea with the \texttt{COLT}\cite{Smith15} radiative transfer code with which we implement a simple shell model (directly analogous to Ly$\alpha$ shell models) for H$\beta$. The dominant parameters in these calculations are the thermal velocities of the inner and outer shell, the relative velocities between the shells, and the optical depth encountered by H$\beta$. 

In Extended Data Fig. 4 we show how a relatively narrow Gaussian H$\beta$ line (arising, say, from a broad-line region) manifests as a double-peaked profile. We confirm that the basic trends seen with the Ly$\alpha$ line hold up for Balmer resonant scattering, and can serve as a powerful guide to interpret the BH*. With this simple toy model we are able to generate a match for the velocity separation and intensity of the strongest peaks seen in the MoM-BH*-1 H$\beta$ profile.

If scattering is underway in the BH*, this has some important implications for the physics of the situation. The success of these simple shell models in producing widely spaced double-peaked structures points to spherical symmetry and a high covering fraction. If the covering fraction were low, there would be flux escaping at line center via ``holes" and tightly spaced peaks as seen in Ly$\alpha$ profiles of Lyman continuum leakers \cite{Rivera-Thorsen17,Vanzella18,NM22}. Furthermore, the relative equality of the peaks means that the structure is more or less at rest (i.e., there is no net velocity between the shells\cite{Verhamme06}) and consistent with negligible dust (since the blue peak photons take a longer path through the scattering medium\cite{Gronke17}). An important implication is that the observed width of Balmer lines (especially when fitted as an absorption system to a broad Gaussian line; i.e. the most common baseline model\cite{Matthee24,Labbe24Monster,deugenio25}) may not trace the kinematics of the broad-line region but are a consequence of resonant scattering. Therefore, SMBH masses in these systems that are based on line-widths of the Balmer lines may be severely overestimated by up to $\approx2$ dex (Extended Data Table 4).

\begin{table}[t]
\centering
\caption{SMBH properties derived using local scaling relations based on H$\beta$\cite{Vestergaard06}. The key caveat is that it is unclear if these relations are applicable given the extraordinary physical conditions observed in this object. See section in text on SMBH properties for details.}
\begin{tabular}{lr}
\hline
\multicolumn{2}{c}{Corrected for absorption and dust (typical assumptions)} \\
\hline
Black Hole Mass [$\log(M_{\rm{BH}}/M_{\rm{\odot}})$] & $8.3^{+0.1}_{-0.2}$
\\
Bolometric luminosity & $45.7^{+0.1}_{-0.1}$\\
Eddington Luminosity [$L/L_{\rm{Edd.}}$] & $0.25^{+0.10}_{-0.04}$\\
\hline
\multicolumn{2}{c}{Corrected for absorption only} \\
\hline
Black Hole Mass [$\log(M_{\rm{BH}}/M_{\rm{\odot}})$] & $7.7^{+0.1}_{-0.2}$
\\
Bolometric luminosity [$\log(L/\rm{erg\ s^{-1}})$] & $45.0^{+0.1}_{-0.1}$\\
Eddington Luminosity [$L/L_{\rm{Edd.}}$] & $0.16^{+0.07}_{-0.03}$\\
\hline
\multicolumn{2}{c}{Motivated by Resonant Balmer Scattering} \\
\multicolumn{2}{c}{(example of complex radiative transfer on Balmer lines)} \\
\hline
Black Hole Mass  [$\log(M_{\rm{BH}}/M_{\rm{\odot}})$] & $\approx6$\\
Bolometric luminosity [$\log(L/\rm{erg\ s^{-1}})$] & $\approx45$\\
Eddington Luminosity [$L/L_{\rm{Edd.}}$] & $\approx5$\\
\hline
\end{tabular}
\end{table}

\noindent
{\bf SMBH Properties and Ramifications for LRD parameters} 
\\
\noindent
Given the unique conditions (e.g., unusual gas density) in this source and the possibility of super-Eddington accretion, we must exercise care in applying the local scaling relations typically used to derive SMBH properties. Indeed, a broader implication of our work is that if BH*s lie at the heart of LRDs, the SMBH properties derived for LRDs may be systematically offset (e.g., the SMBH mass is likely overestimated, and the Eddington ratio underestimated). Here we explore implications for a variety of approaches to back out the SMBH properties that are summarized in Table 4. Given the significant systematic uncertainties, these calculations must be seen as order of magnitude estimates to bracket the range of possibilities. 

First, we simply apply H$\beta$-based local scaling relations\cite{Vestergaard06} to the absorption-corrected and dust-corrected H$\beta$ line. The absorption correction is typically done via emission line fitting assuming an underlying Gaussian or Lorentzian\cite{Matthee24,Labbe24Monster} (as in Extended Data Fig. 1). For the dust correction, note that the H$\alpha$ and H$\beta$ line ratio\cite{Furtak24} is challenging to use given that their observed fluxes reflect radiative transfer in the dense gas. The standard approaches used in the literature at the moment infer a significant $A_{\rm{V}}$ of few magnitudes either from continuum slope fitting or through SED modeling\cite{Furtak24,Greene24,Ji25BlackThunder,Brooks24}. Going by the optical continuum slope\cite{Greene24} would imply an $A_{\rm{V}}\approx2$ for MoM-BH*-1 implying a black hole mass of $\approx10^{8.3} M_{\rm{\odot}}$. This is comparable to the stellar mass of the host galaxy ($<10^{8.5} M_{\rm{\odot}}$ at $95\%$ confidence), i.e., this is an apparently ``overmassive" black hole relative to local scaling relations between host galaxy mass and black hole mass\cite{Furtak24,Matthee24ALTclustering}.

However, a key insight from the BH* SED is that the $A_{\rm{V}}$ required to explain the LRDs is likely negligible as the BH* SED is intrinsically UV-weak below the Balmer break. That is, the Little Red Dots are red not due to obscuration from dust, but due to opacity from gas. Accounting for this, with $A_{\rm{V}}=0$, the SMBH mass derived now is $\approx5\times10^{7}M_{\rm{\odot}}$ with a luminosity of $\approx15\%$ the Eddington limit.

Finally, we note that the observed line width (after correcting for absorbers) may not be faithfully tracing the broad-line region, violating the basic ansatz for scaling relations. The underlying BH* emission line profile in LRDs may be as complex as the one shown in Extended Data Fig. 2, but is perhaps difficult to disentangle from strong emission from the host. The rich and symmetric structure may signify complex radiative processes that are modifying the emission arising from the broad-line region. As an example of one such process, if H$\beta$ is undergoing resonant scattering through the atmosphere around the BH*, then the intrinsic BLR width before scattering could be as low as $\approx600$ km $s^{-1}$ (see Extended Fig. 4). This would then yield an SMBH mass of $\approx10^{6} M_{\rm{\odot}}$.

To summarize, there is no precedent for this source, and so it is challenging to estimate the SMBH properties -- but importantly, these challenges may extend to the entire LRD population. We have presented a range of approaches from local scaling relations to tailored estimates based on our best guess for the physics of this object. There is an up to 2.5 dex difference in the SMBH mass estimated across these approaches, signaling that the SMBH masses reported for LRDs may be systematically overestimated. The stellar mass to black hole mass ratios reported for these sources must be treated with caution.

\noindent
{\bf Host Galaxy Properties -- Are LRDs quenching their dwarf galaxy hosts?}
\\
\noindent
To constrain the host galaxy mass, we can leverage the recent insight from clustering analyses that the rest-UV light in the typical LRD (FWHM=$1000-2000$ km s$^{-1}$), on average, originates almost entirely from the host galaxy\cite{Matthee24ALTclustering}. Of course, this is not true for all LRDs, particularly the most luminous sources with higher FWHM broad lines which display AGN signatures even in the rest-UV\cite{Labbe24Monster}, but so far they seem the exception. The UV luminosity therefore still allows for an interesting upper limit on the host mass. To construct an empirical $M_{\rm{\star}}-M_{\rm{UV}}$ relation we use the compilation of low-luminosity galaxies at $z=3-7$ from the All the Little Things (ALT) survey\cite{Naidu24} in the Abell 2744 field. This is the largest spectroscopic sample of $M_{\rm{UV}}<-15$ galaxies at these redshifts. The ALT stellar masses are derived with the \texttt{prospector} SED fitting code applied to 27 bands of NIRCam+HST photometry, including all JWST medium and broad bands\cite{Suess24,Bezanson22}. From this sample we estimate that a source with $M_{\rm{UV}}$ between $-17.9$ and $-18.3$ has a $95\%$ upper limit on its stellar mass of $\log(M_{\rm{\star}}/M_{\rm{\odot}})<8.5$.

\begin{figure*}[h!]
  \centering
\includegraphics[width=\textwidth]{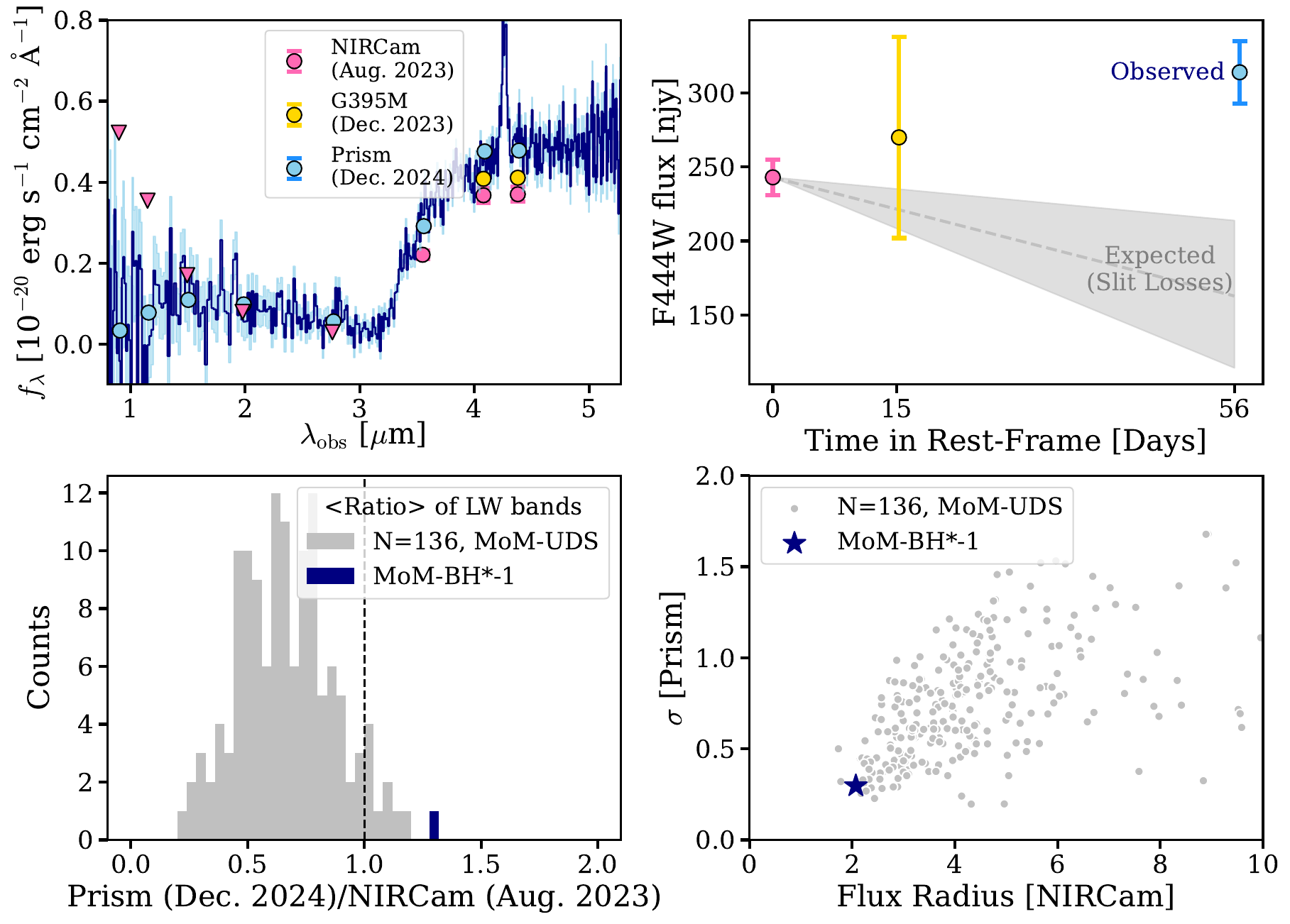}
  \caption{\textbf{Extended Data Figure 5 | Testing variability in MoM-BH*-1.} \textbf{Panel a.} Synthesized NIRCam photometry from the three epochs of observations are compared (pink, gold, blue) against the prism spectrum (navy). Non-detections in the NIRCam imaging (2$\sigma$ upper limits) are shown as pink triangles. Note that the spectrum is generally deeper than the photometry, particularly in the rest-UV. \textbf{Panel b.} The observations span 56 days in the rest-frame of the source. While virtually every source in the prism observations has lost flux relative to the F444W imaging due to slit-losses (1$\sigma$ contours for median ratio shown in gray), MoM-BH*-1 has brightened by $\approx30\%$. \textbf{Panel c.} MoM-BH*-1 is the only source across both observed masks in the UDS that has brightened by this degree relative to its NIRCam photometry. The ratio shown here is the median ratio of the flux synthesized from the prism and directly measured by NIRCam in the four LW filters observed by the PRIMER survey in August 2023 (F277W, F356W, F410M, and F444W). \textbf{Panel d.} This variability signal is unlikely to be due to an overestimated extraction aperture ($\sigma$) in the NIRSpec reduction pipeline. The compact flux radius of the source measured from the imaging is commensurate with the extraction aperture used for the spectra.}. 
  \vspace{-4mm}
\end{figure*}

Another estimate of the stellar mass and detailed properties of the host can be derived via the SED subtraction technique shown in Fig. 4 which assumes all the UV light arises from the host galaxy (again, providing an upper limit). In particular, attributing the narrow [OIII] flux to the host galaxy while requiring it to lie at $M_{\rm{UV}}>-18.1$ with a consistent $\beta_{\rm{UV}}$ slope strongly constrains the allowed SED shape. We empirically constrain this shape with a stack of 53 sources selected from the DJA with appropriate UV slopes and low [OIII] fluxes. Similar to the host SED shown in Fig. 4, this stack displays a Balmer break along with weak [OIII] emission -- given its low luminosity, this is likely a galaxy with relatively little recent star-formation\cite{Strait23,Witten24,Looser24}.We fit its properties using the \texttt{Prospector} SED fitting code and estimate a stellar mass limit of $\log{M/M_{\rm{\star}}}<8.9$ consistent with the earlier calculation.

\noindent
{\bf Hints of Variability -- $4\sigma$ detection, but across different instruments}
\\
\noindent
It is of particular interest to test for signs of variability in MoM-BH*-1 as an independent constraint on the physics of the source. AGN display variability on all timescales\cite{Ulrich97,MacLeod10,Shen24SRM}. Furthermore, the remarkable gas conditions of this source makes it a particularly promising candidate for monitoring. Super-Eddington accretion and the resulting instabilities, inhomogeneities, and holes in the ``atmosphere" of the BH* might produce dramatic variability. In LRDs, the SED is a summation of the host and BH* at all wavelengths to varying degrees, but in this case, the stark outshining of the host galaxy means an ``undiluted" variability signal may be stronger and more easily detected. 

As discussed above, three sets of observations covering $3-5\mu$m exist for MoM-BH*-1 separated by $\approx60$ days in the rest-frame from the PRIMER (NIRCam), EXCELS (NIRSpec/G395M), and MoM (NIRSpec/prism) surveys. These observations are illustrated in Extended Data Figure 2. It is certainly not ideal to test for variability across three different observing modes with distinct systematics. Nevertheless, it is notable that the source appears to have brightened by $30\pm7\%$ between the first epoch (NIRCam) and third epoch (NIRSpec prism, no post-processing renormalization to photometry). The prism spectrum suffers from slit losses, and yet this source appears brighter than expected from the NIRCam photometry. It is the only source across the 136 sources with high-SNR photometry (SN$>10$) and spectra (50$^{\rm{th}}$ percentile of SN$>5$) observed in two masks as part of the MoM program that shows this degree of brightening. We also note that typically, the NIRSpec/G395M flux is systematically $\approx10-20\%$ lower than the prism flux due to calibration uncertainties\cite{deGraaff24RUBIES} -- accounting for this, both the prism and NIRSpec fluxes for MoM-BH*-1 appear to be in agreement and $\approx30\%$ brighter than the NIRCam flux. Note that the shape of the SED (and the depth of the Balmer break) is consistent across the NIRCam and prism data (e.g., $f^{\rm{\nu}}_{\rm{F410M}}/f^{\rm{\nu}}_{\rm{F356W}}=2.0\pm0.2$ in the prism vs. $2.2\pm0.2$ in NIRCam). The difference is in the absolute brightness.

Taken at face value, these measurements imply a $\approx4\sigma$ detection of variability at rest-optical wavelengths over a mere two months. This may be independent evidence not only for the AGN nature of the source, but also for the volatile environment that exists around the BH*. This remarkable degree of brightening marks MoM-BH*-1 as an excellent target for future monitoring campaigns. 

\begin{addendum}
  
\item [Acknowledgements] We thank the PRIMER and EXCELS teams for designing JWST surveys that made this work possible. We acknowledge feedback on "BH*" and illuminating discussions with Ivan Cabrera-Ziri, Mike McDonald, Roger Blandford, Susan Clark, Risa Wechsler, Tom Abel, Mike Liu, Dave Sanders, Sylvain Veilleux, Triana Almeyda, Adam Ginsburg, and Desika Narayanan. We thank Kohei Inayoshi for discussing details of their work modeling Balmer breaks in LRDs\cite{Inayoshi24}. 
This work is based on observations made with the NASA/ESA/CSA James Webb Space Telescope. The data were obtained from the Mikulski Archive for Space Telescopes at the Space Telescope Science Institute, which is operated by the Association of Universities for Research in Astronomy, Inc., under NASA contract NAS 5-03127 for JWST. These observations are associated with programs 5224 and 3543.
Some of the data products presented herein were retrieved from the Dawn JWST Archive (DJA). DJA is an initiative of the Cosmic Dawn Center (DAWN), which is funded by the Danish National Research Foundation under grant DNRF140.
We acknowledge funding from JWST programs GO-3516, GO-5224, and GO-1837. Support for this work was provided by NASA through the NASA Hubble Fellowship grant HST-HF2-51515.001-A awarded by the Space Telescope Science Institute, which is operated by the Association of Universities for Research in Astronomy, Incorporated, under NASA contract NAS5-26555. Funded by the European Union (ERC AGENTS, 101076224; HEAVYMETAL, 101071865; RED CARDINAL, 101076080). Views and opinions expressed are however those of the author(s) only and do not necessarily reflect those of the European Union or the European Research Council. Neither the European Union nor the granting authority can be held responsible for them. This work has received funding from the Swiss State Secretariat for Education, Research and Innovation (SERI) under contract number MB22.00072, as well as from the Swiss National Science Foundation (SNSF) through project grant 200020\_207349. This work was also supported by JSPS KAKENHI Grant Numbers 23H00131. The Cosmic Dawn Center is funded by the Danish National Research Foundation
under grant DNRF140. P.N. acknowledges support from the Gordon and Betty Moore Foundation and the John Templeton Foundation that fund the Black Hole Initiative (BHI) at Harvard University where she serves as an external PI. S.B. acknowledges funding from a UK Research \& Innovation (UKRI) Future Leaders Fellowship [grant number MR/V023381/1]. 

\item[Author Contributions] All authors contributed to aspects of the analysis and to the writing of the manuscript.

\item[Author Information] Correspondence and requests for materials should be addressed to RPN (rnaidu@mit.edu).

\item[Data Availability] The prism spectra obtained as part of JWST-GO-5224 (``Mirage or Miracle") featured in this work are available on Zenodo. All processed data from this program will eventually be incorporated in the DAWN JWST archive. All other processed images and spectra used in this work are publicly available via the DAWN JWST archive (\url{https://dawn-cph.github.io/dja/}). 

\item[Code Availability] All results presented may be reproduced with the open access reduced data described above and using the following publicly available software: \texttt{msaexp}, \texttt{grizli}, \texttt{astropy}, \texttt{Cloudy},
\texttt{SpectRes}, \texttt{pysersic}, \texttt{COLT}, \texttt{numpyro}.

\end{addendum}

\bibliography{masterbiblio}

\let\thefootnote\relax\footnote{

\begin{affiliations}
\item MIT Kavli Institute for Astrophysics and Space Research, 70 Vassar Street, Cambridge, MA 02139, USA
\item NASA Hubble Fellow
\item Institute of Science and Technology Austria (ISTA), Am Campus 1, 3400 Klosterneuburg, Austria
\item Department of Astronomy \& Astrophysics, University of Chicago, 5640 S Ellis Avenue, Chicago, IL 60637, USA
\item Max-Planck-Institut f"ur Astronomie, K"onigstuhl 17, D-69117 Heidelberg, Germany
\item Department of Astronomy, University of Geneva, Chemin Pegasi 51, 1290 Versoix, Switzerland
\item Cosmic Dawn Center (DAWN), Copenhagen, Denmark
\item Niels Bohr Institute, University of Copenhagen, Jagtvej 128, Copenhagen, Denmark
\item Department of Physics, The University of Texas at Dallas, Richardson, Texas 75080, USA
\item Department of Astrophysical Sciences, Princeton University, Princeton, NJ 08544, USA
\item Department of Astronomy, The University of Texas at Austin, Austin, TX 78712, USA
\item Centre for Astrophysics and Supercomputing, Swinburne University of Technology, Melbourne, VIC 3122, Australia
\item Institut d'Astrophysique de Paris, CNRS, Sorbonne Universit'e, 98bis Boulevard Arago, 75014, Paris, France
\item Department of Astronomy, Yale University, New Haven, CT 06511, USA
\item Dipartimento di Fisica e Astronomia, Università di Bologna, Bologna, Italy
\item Department of Physics and Astronomy and PITT PACC, University of Pittsburgh, Pittsburgh, PA 15260, USA
\item Leiden Observatory, Leiden University, PO Box 9513, 2300 RA Leiden, The Netherlands
\item Institute for Computational Cosmology, Department of Physics, Durham University, South Road, Durham, DH1 3LE, UK
\item Kapteyn Astronomical Institute, University of Groningen, P.O. Box 800, 9700 AV Groningen, The Netherlands
\item Center for Frontier Science, Chiba University, 1-33 Yayoi-cho, Inage-ku, Chiba 263-8522, Japan
\item Department of Physics, Ben-Gurion University of the Negev, P.O. Box 653, Be'er-Sheva 84105, Israel
\item Max Planck Institute for Astrophysics, Karl-Schwarzschild-Str. 1, 85748 Garching, Germany
\item Institute of Physics, Lab for galaxy evolution and spectral modelling, EPFL, Observatory of Sauverny, Chemin Pegasi 51, 1290 Versoix, Switzerland
\item Department of Astronomy and Astrophysics, University of California, Santa Cruz, CA 95064, USA
\item Department of Physics, School of Advanced Science and Engineering, Faculty of Science and Engineering, Waseda University, 3-4-1 Okubo, Shinjuku, Tokyo 169-8555, Japan
\item Waseda Research Institute for Science and Engineering, Faculty of Science and Engineering, Waseda University, 3-4-1 Okubo, Shinjuku, Tokyo 169-8555, Japan
\item Center for Astrophysics | Harvard and Smithsonian, 60 Garden Street, Cambridge, MA 02138, USA
\item Department of Astronomy \& Astrophysics, The Pennsylvania State University, University Park, PA 16802, USA
\item Institute for Computational \& Data Sciences, The Pennsylvania State University, University Park, PA 16802, USA
\item Institute for Gravitation and the Cosmos, The Pennsylvania State University, University Park, PA 16802, USA
\item GRAPPA, Anton Pannekoek Institute for Astronomy and Institute of High-Energy Physics, University of Amsterdam, Science Park 904, NL-1098 XH Amsterdam, the Netherlands
\item Department of Astronomy, University of Wisconsin-Madison, 475 N. Charter St., Madison, WI 53706 USA
\item Department of Physics, Yale University, New Haven, CT 06511, USA
\item Yale Center for Astronomy \& Astrophysics, Yale University, New Haven, CT 06520, USA
\item Department for Astrophysical and Planetary Science, University of Colorado, Boulder, CO 80309, USA
\item Brinson Prize Fellow
\item Centro de Astrobiolog{'i}a (CAB), CSIC-INTA, Ctra. de Ajalvir km 4, Torrejón de Ardoz, E-28850, Madrid, Spain
\item BNP Paribas Corporate \& Institutional Banking, Torre Ocidente Rua Galileu Galilei, 1500-392 Lisbon, Portugal
\item Departamento de F'isica, Faculdade de Ciencias, Universidade de Lisboa, Edif'icio C8, Campo Grande, PT1749-016 Lisbon, Portugal
\item Departament d'Astronomia i Astrof`isica, Universitat de Val`encia, C. Dr. Moliner 50, E-46100 Burjassot, Val`encia, Spain
\item Unidad Asociada CSIC ``Grupo de Astrof'isica Extragal'actica y Cosmolog'ia" (Instituto de F'isica de Cantabria - Universitat de Val`encia)
\item Kavli Institute for Cosmology, University of Cambridge, Madingley Road, Cambridge, CB3 OHA, UK
\item Cavendish Laboratory, University of Cambridge, 19 JJ Thomson Avenue, Cambridge CB3 0HE, UK
\item Sterrenkundig Observatorium, Universiteit Gent, Krijgslaan 281 S9, 9000 Gent, Belgium
\end{affiliations}
}

\end{document}